\documentclass[%
 reprint,
superscriptaddress,
 amsmath,amssymb,
 prl,
nofootinbib
]{revtex4-2}

\usepackage{graphicx}
\usepackage{dcolumn}
\usepackage{bm}

\begin{document}


\title{
Balanced and fragmented phases in societies with homophily and social balance}

\author{Tuan Minh Pham}
\affiliation{Section  for  the  Science  of  Complex  Systems,  CeMSIIS, Medical  University  of  Vienna,  Spitalgasse  23,  A-1090,  Vienna,  Austria
}
\affiliation{Complexity  Science  Hub,  Vienna  Josefst\"{a}dterstrasse  39,  A-1090  Vienna,  Austria}

\author{Andrew C. Alexander}
 \affiliation{Department of Mathematics, Princeton University, NJ 08544, Princeton,  USA
 }

\author{Jan Korbel}
\affiliation{Section  for  the  Science  of  Complex  Systems,  CeMSIIS, Medical  University  of  Vienna,  Spitalgasse  23,  A-1090,  Vienna,  Austria
}
\affiliation{Complexity  Science  Hub,  Vienna  Josefst\"{a}dterstrasse  39,  A-1090  Vienna,  Austria}

\author{Rudolf Hanel}
\affiliation{Section  for  the  Science  of  Complex  Systems,  CeMSIIS, Medical  University  of  Vienna,  Spitalgasse  23,  A-1090,  Vienna,  Austria
}
\affiliation{Complexity  Science  Hub,  Vienna  Josefst\"{a}dterstrasse  39,  A-1090  Vienna,  Austria}

\author{Stefan Thurner}
\affiliation{Section  for  the  Science  of  Complex  Systems,  CeMSIIS, Medical  University  of  Vienna,  Spitalgasse  23,  A-1090,  Vienna,  Austria
}
\affiliation{Complexity  Science  Hub,  Vienna  Josefst\"{a}dterstrasse  39,  A-1090  Vienna,  Austria}
\affiliation{Santa  Fe  Institute,  1399  Hyde  Park  Road,  Santa  Fe,  NM  87501,  USA}

\date{\today}

\begin{abstract}
Recent attempts to understand the origin of social fragmentation are based on spin models which include terms accounting for two social phenomena: homophily---the tendency for people with similar opinions to establish positive relations---and social balance---the tendency for people to establish balanced triadic relations. Spins represent attribute vectors that encode multiple (binary) opinions of individuals and social interactions between individuals can be positive or negative. 
Recent work suggests that large systems of   $N \gg 1$ individuals never reach a balanced state (where unbalanced  triads with one or three hostile links remain),  provided the number of attributes for each agent is less than $O(N^2)$ [Phys. Rev. Lett. 125, 078302]. 
Here we show that this statement is overly restrictive. 
Within a Hamiltonian framework that minimizes individuals' social stress, we demonstrate that stationary, balanced, but  \emph{fragmented} states can be reached  for \textit{any} number of attributes, if, in addition to homophily, individuals take into account a significant fraction,  $q$,  of their triadic relations. Above a critical value $q_c$, balanced states  result. 
 This result also holds for sparse realistic social networks. Finally, in the limit of small $q$, our result agrees with that of [Phys. Rev. Lett. 125, 078302].
\end{abstract}

\maketitle

The concept of so-called {\em filter bubbles} captures the fragmentation of society into isolated groups of people who trust each other, but clearly distinguish themselves from ``other''. Opinions tend to align within groups and diverge between them. 
 Interest in this process of social disintegration, started by Durkheim \cite{Durkheim},  has experienced a recent boost, fuelled by the availability of modern communication technologies. The extent to which societies fragment depends largely on the interplay of two basic mechanisms that drive social interactions:  \emph{homophily}  and \emph{structural balance}. Homophily is based on the ``principle'' that individuals with similar opinions tend to become friends (``similarity breeds connection''  \cite{Cook}); indeed, like-minded individuals often form homogeneous structures in society \cite{Huber}. \emph{Structural balance}, first described by Heider \cite{Heider}, is the tendency of balanced triads to be over-represented in societies. A triad of individuals is \emph{balanced} if all three individuals are mutual friends 
({friend of my friend is my friend}) or if two friends have a mutual enemy ({enemy of my enemy is my friend}). 
Structural balance has been investigated by social scientists for a long time \cite{Harary, cartwright1956, David} and, more recently, by physicists and network scientists \cite{Kulakowski, Antal2005, Antal2006, Radicchi, Marvel2010, Traag2013,  Hummon, Abell2009, Szell,Leskovec, sadilek}. Recent contributions study the dynamics on balanced networks \cite{Altafini, Proskurnikov, Cisneros-Velarde}, the co-evolution of opinions and signed networks  \cite{Baumann, Gross2019, Gao2018, Chen2014, DENG2016, Singh, Singh2016, Macy,Antonio,Aguiar,Agbanusi, Saeedian2019, Saeedian2020, Tuan},
and generalizations of the concept of structural balance \cite{Kirkley}. For an overview, see \cite{Zheng, Rawlings}. A general survey of statistical physics methods applied to opinion dynamics is found in \cite{Castellano,Galesic}.

Among the various approaches towards social fragmentation, the minimization of social stress captured by a so-called \emph{social stress Hamiltonian} has become particularly relevant. Here the opinion of individual $i$ is denoted by $s_i$ and the relation between $i$ and $j$ by $J_{ij}$  (positive or negative). Stress arises from a homophily-related term, $- \sum_{(i,j)} J_{ij} s_i s_j$, or a term reflecting social balance, $-\sum_{(i,j,k)} J_{ij}J_{jk}J_{ki}$. The former is similar in structure to the Edwards-Anderson spin-glass model \cite{Edwards1975},  the latter  to the Baxter-Wu three-spin interaction model \cite{Baxter}. These two terms   have been studied  separately, homophily in  \cite{AB1993, Galam1996, Vinogradova2013,Facchetti2011}  and Heider balance in \cite{Marvel, Rabbani, Belaza2017}; only recently were both terms  integrated into a single Hamiltonian \cite{Tuan}. 

The issue of social cohesion and organization was recently studied in an attribute-based local triad dynamics model (ABLTD) \cite{Gorski}. There, each agent has binary opinions on $G$ attributes. If two agents agree on more attributes than they disagree on, they become friends (positive link). 
Agents tend to change their attributes to reduce stress in unbalanced triads. 
The paper showed that given a system of $N$ agents, as $N \to \infty$, the so-called ``paradise state'', 
 where all agents are friends of each other, 
is never reached unless the number of attributes $G$ scales as $O(N^\gamma)$, for $\gamma \geq 2$.  
Instead, the society remains in a stationary \emph{unbalanced} state with an equal number of balanced and unbalanced triads. This means that for realistic situations, where $N$ is large and the number of attributes $G$ remains  relatively small, 
it is impossible to reach social balance, let alone the paradise state.  
This  statement is to some extent, contrary to empirical findings that societies are balanced to a high degree; see e.g. recent work on large scale studies \cite{Szell,Leskovec}. 

In this letter, motivated by the ABLTD, we propose  an \emph{individual-stress-based} model that takes into account the homophily effect between adjacent individuals and structural balance within a local neighborhood. The latter consists of the subset of  the most relevant triads to an individual, i.e. those that involve its closest friends and greatest enemies. 
The ratio of relevant triads to the total number of triangles the individuals belong to determines whether society fragments or remains cohesive.
With the help of simulations on a regular network, we show that there exists a critical size of the local neighborhood above which society  fragments,  yet  stays balanced.
We discuss the relation of the presented model to both, the ABLDT model \cite{Gorski} and the social stress Hamiltonian approach \cite{Tuan}. For appropriate parameter choices, both models can be shown to correspond to special cases of the presented model.

\paragraph*{Local social stress model.}
Consider a society of $N$ individuals. Each individual $i$ has binary opinions on $G$ issues, characterized by an attribute vector, $\textbf{A}_i = \{a_i^{\ell}\}$, where $a_i^{\ell} \in \{-1,+1\}$; $\ell \in 1,\dots,G$.  Further, $i$ has relations to $k_i$ other individuals in a social network. Network   topology does not change over time.  
Following \cite{Gorski}, the relation between two agents $i$ and $j$  is determined by the sign of their distance in attribute space:  $J_{ij} = {\rm sign}(\textbf{A}_i \cdot \textbf{A}_j)$,  where the dot denotes the scalar product.  $J_{ij} = 1$ indicates friendship, $J_{ij}=-1$  enmity. Each agent $i$ has a social  stress level, $H^{(i)}$, defined as
\begin{equation}
H^{(i)}(\textbf{A})= -   \frac{1}{G}\,  \sum_{j} J_{ij} \textbf{A}_i \cdot \textbf{A}_j -\sum_{(j,k)_{Q_{i}}} J_{ij}J_{jk}J_{ki} \, .
 \label{local}
\end{equation}
The first sum extends over all $k_i$ neighbours of $i$, while  the second is restricted to $Q_i$ (out of $N^{\Delta}_i$ possible\footnote{By definition, $N^{\Delta}_i \equiv  c_i k_i(k_i-1)/2$, where $ c_i $ is the local clustering coefficient. $k_i$ is the degree.}) triads that node $i$ belongs to. The notation $(j,k)_{Q_i}$ means to sum over all pairs of $j$ and $k$ which, together with $i$, form 
the $Q_i$  triads. These are chosen at each step of the dynamics (see below).
$Q_i$ represents the number of triads $i$ would like to have socially balanced  -- $i$ 's relevant neighborhood. 
The factor $1/G$ ensures that contributions from any link towards $H^{(i)}$ do not diverge in the limit $G \rightarrow \infty$.  Assuming agents try to minimize their individual social stress over time, we implement the following dynamics\footnote{The relevance of the first term in the model dynamics is discussed in the Supplemental Material.}: 

\begin{figure}
\includegraphics[scale=0.23]{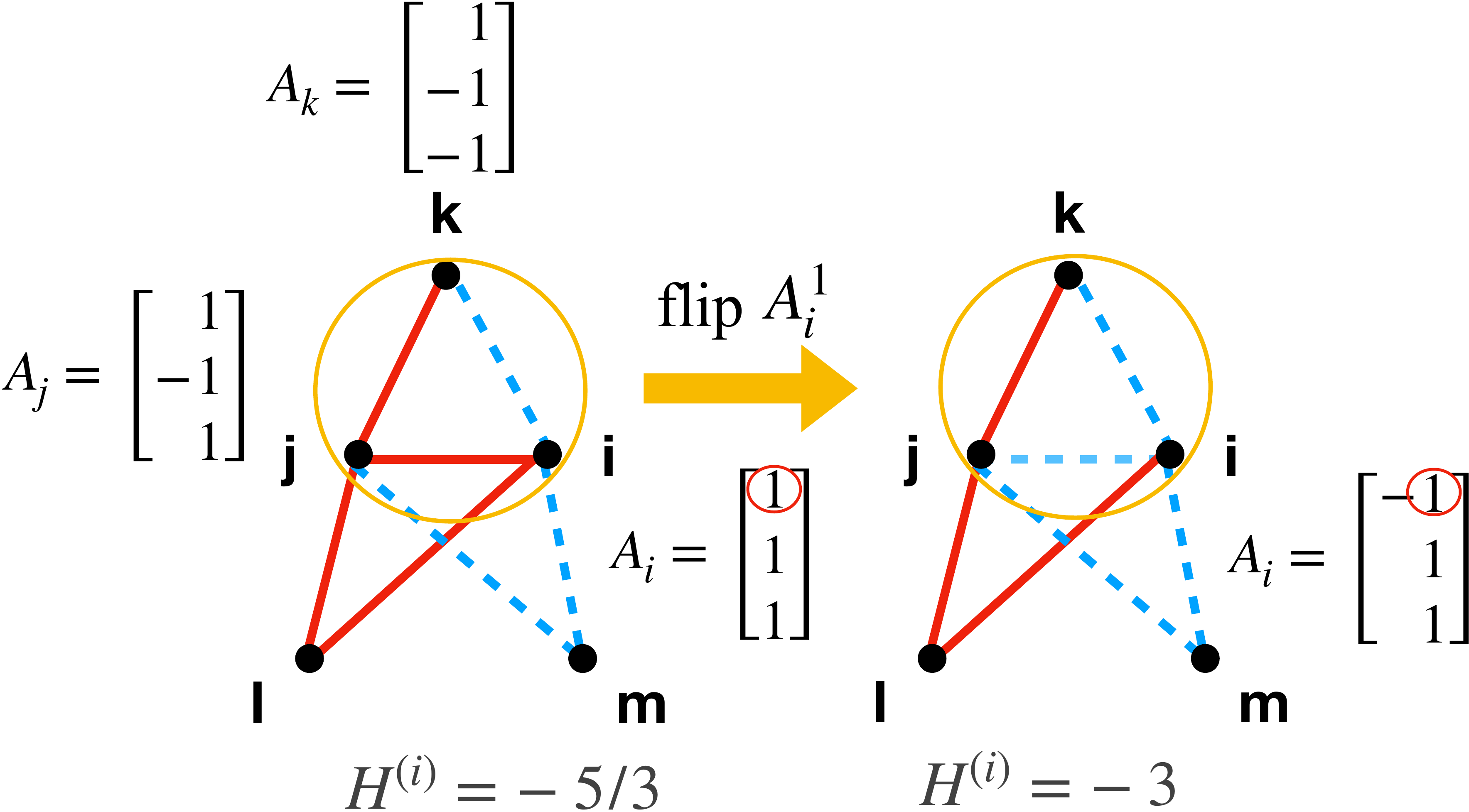}
\caption{Co-evolutionary interplay of opinions and links. Red (blue) links denote positive (negative) relationships.  Among the three triads in the graph, the chosen  one, $(i, j , k)$ is circled. We consider the case $Q = 1$.  As agent $i$ flips one of its attributes, $A_i^1$, this triad becomes balanced and $i$ decreases its individual stress from $-5/3$ to $-3$. The  opinion vectors of $l$ and $m$ are $\textbf{A}_l = \textbf{A}_i$ and $\textbf{A}_m = - \textbf{A}_i$, respectively (not shown in the figure). For $Q = 2$,  the same flip of  $A_i^1$ leads to an increase of $\Delta H^{(i)}_{Q = 2} = 2/3$, if one of the two chosen triads contains $k$, or $\Delta H^{(i)}_{Q = 2} = 14/3$, if none of them contains $k$. For $Q = 3$,  $\Delta H^{(i)}_{Q = 3} =   8/3$. The higher $Q$, the more important becomes the role of social balance.
}
\label{fig:step}
\end{figure}

\begin{enumerate}
\item \emph{Initialize}.  Each node is assigned  an opinion vector, $\textbf{A}_i$, whose components are randomly chosen to be $ 1$ or $-1$ with equal probability. Every node has the same degree, $k_i = K$, and  is connected to its neighbours in a regular way, forming a  ring topology. The topology is fixed over time. 
For any pair of connected agents, $i$ and $j$, we set $J_{ij} = {\rm sign}(\textbf{A}_i \cdot \textbf{A}_j)$. 
\item \emph{Update}.
{\em (i)} Pick a node $i$ randomly and choose $Q$ of its triads, also randomly. Compute $H^{(i)}$. In the current state its  value is $\mathcal{H}$. 

{\em (ii)}  Flip one of $i$'s attributes at random. Let $\tilde{\textbf{A}}_i$ be its new opinion vector. For each of the chosen triads, the weights of the two links adjacent to $i$ are recomputed  as $\tilde{J}_{ij} = {\rm sign}(\tilde{\textbf{A}}_i \cdot \textbf{A}_j)$.  $\tilde{J}$ is the new matrix. 

{\em (iii)}   Compute the new stress $\tilde{\mathcal{H}}$ using $\tilde{J}$. The change in stress is $\Delta H^{(i)} \equiv \tilde{\mathcal{H}}-\mathcal{H}$.

{\em (iv)}  Update the system $\textbf{A}_i \rightarrow \tilde{\textbf{A}}_i$ and $J_{ij} \rightarrow \tilde{J}_{ij}$ with probability,   $\min\left\{e^{ - \Delta H^{(i)}},1\right\}$, otherwise leave it unchanged. This stochastic rule means that agents are not always rational and might choose to increase their stress.

\item Continue with the next timestep.
\end{enumerate}
 
Figure \ref{fig:step} illustrates an update where by changing one opinion, agent $i$  becomes an enemy of $j$, but the chosen triad, $(ijk)$, becomes balanced. If $Q=1$, this decreases $i$'s social stress from $H^{(i)} = -5/3$ to $H^{(i)} = -3$.  
If more triads are chosen, $Q>1$, this flip leads to a stress increase and is less likely accepted.

According to the dynamical rule, 
the maximum number of links which may change their signs due to an opinion update depends on $Q$. When $Q = 1$, at most  two links can flip, while for $Q \geq K$  all the $K$ links might. 
The change in stress for agent $i$, $\Delta H^{(i)}$, given that attribute $a_i^{\ell_*}$ flips,  can be written as 
\begin{equation}
\Delta H^{(i)}=  \frac{2}{G}\, \sum_{(j | \tilde{J}_{ij} = J_{ij})} J_{ij}  a_i^{\ell_*}  a_j^{\ell_*}  +  \sum_{(j,k)_Q}  \Delta_{jk} \,  ,
\label{localchange}
\end{equation}
where $j | \tilde{J}_{ij} = J_{ij}$ means to sum over those $j$ (neighbours of $i$) whose  relations to $i$ remain unchanged and $\Delta_{jk} = \big[J_{ij}J_{ki} - \tilde{J}_{ij}\tilde{J}_{ki}\big] J_{jk} $. Obviously, $\Delta_{jk} \in \{-2, 0, 2\}$.  
Since most links are kept frozen for small $Q$, 
a flip in  $a_i^{\ell_*}$  is more likely to be accepted if it increases the concordance in opinions between $i$ and its friends, and likewise the  discordance in opinions between $i$ and its enemies. This kind of dynamics drives friends to become more similar while enemies drift  apart. Note the similarity to the Hebbian rule \cite{Hebb, Macy}. 
Because of the random assignment of the opinions at the start,  there are approximately as many balanced as unbalanced triads in the stationary state. 
For large $Q$, the question is whether  there  exists a critical $Q_c$ above which social balance dominates homophily. If such $Q_c$ exists, the social network may reach a balanced state without frustrated triads. Once  global balance is established, the 
opinion dynamics  converges to states where all individuals have a minimum amount of social stress.

Both cases, small and large $Q$, are affected by the number of attributes, $G$. The probability that a link incident with $i$ switches its sign if $a_i^{\ell_*}$  flips, is proportional to $1/\sqrt{G}$, as $G \rightarrow \infty$. Thus large $G$ can  prevent the system from reaching a balanced state because links are less likely to change. On a sparse network, the larger $G$ is, the higher $Q$ must be to reach social balance. Finally, if one keeps $q = Q/N^{\Delta}$ fixed, then as $N^{\Delta}$ grows with $K$, $Q$ must also grow with $K$, making balanced states more likely to be achieved for high $K$.

\begin{figure}[tb]
\includegraphics[width=\linewidth]{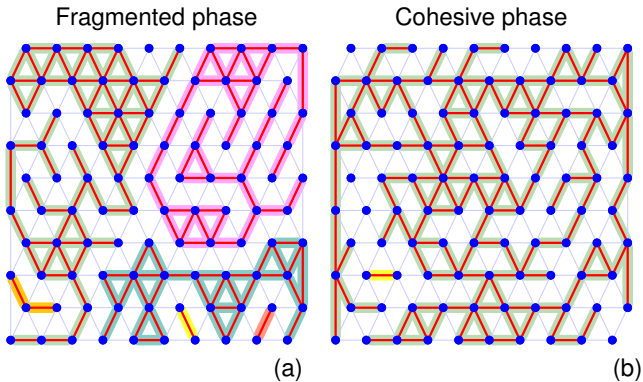}
\caption{(a) Clusters of positive links (red) are identified in the fragmented and balanced phase, characterized by $G = 3$ opinions, while in (b) a single cluster appears in the cohesive phase for $G = 99$. $N= 100$, $Q = 2$. Note that $N^{\Delta} = 2, 3$, or $6$, depending on the position of the agents on the lattice (free boundary conditions). Clusters are highlighted with colors.}
\label{fig:2}
\end{figure}

\paragraph*{Order parameter:}
To measure the level of social balance within a society, we define an order parameter, $f$,  as the difference between the proportions of balanced and unbalanced triangles: 
 \begin{equation}
 f = \frac{n_{+}  - n_{-}}{n_{+} + n_{-}} \, ,
 \label{f}
 \end{equation}
where  $n_{+}$ and $n_{-}$ are the  number of  balanced and unbalanced triangles, respectively.  
$f= 1$ means that all triangles are balanced, $f < 1$ signals the presence of unbalanced triangles. Even though $f$ could be negative, this situation is never observed in simulations.
A fundamental result in structural balance theory states that if the network can be partitioned into  clusters, within which all links are positive and between which links are exclusively negative, then all triangles are balanced, and $f = 1$ \cite{Wasserman}. Note that $ f = 1$ is  a necessary, but not  sufficient condition for such a partition on sparse networks.  We therefore call a society \emph{balanced and fragmented} if  all of its constituent triads are balanced and clusters of positive links are separated from each other by negative links.

\paragraph*{Results.}
We first run the dynamics on a triangular lattice for $N=100$ with free boundary conditions to check whether clusters of positive links can be formed. To this end, we set $Q = 2$ -- i.e. only two triads per individual are chosen in each update. In Fig. \ref{fig:2}  we show a snapshot of the signed social network after reaching a stationary state. For a small value of $G = 3$, the formation of several clusters of positive (red)  links is clearly seen. Negative links are shown in light blue. This configuration belongs to the \emph{fragmented phase} that shows a high degree of balance, $f \sim 0.9$. For more attributes, $G = 99$,  a single cluster emerges that percolates the lattice. This is a realization in the \emph{cohesive phase} that is clearly not balanced, and indeed  $f \sim 0$.

\begin{figure}[tb]
\includegraphics[scale=0.255]{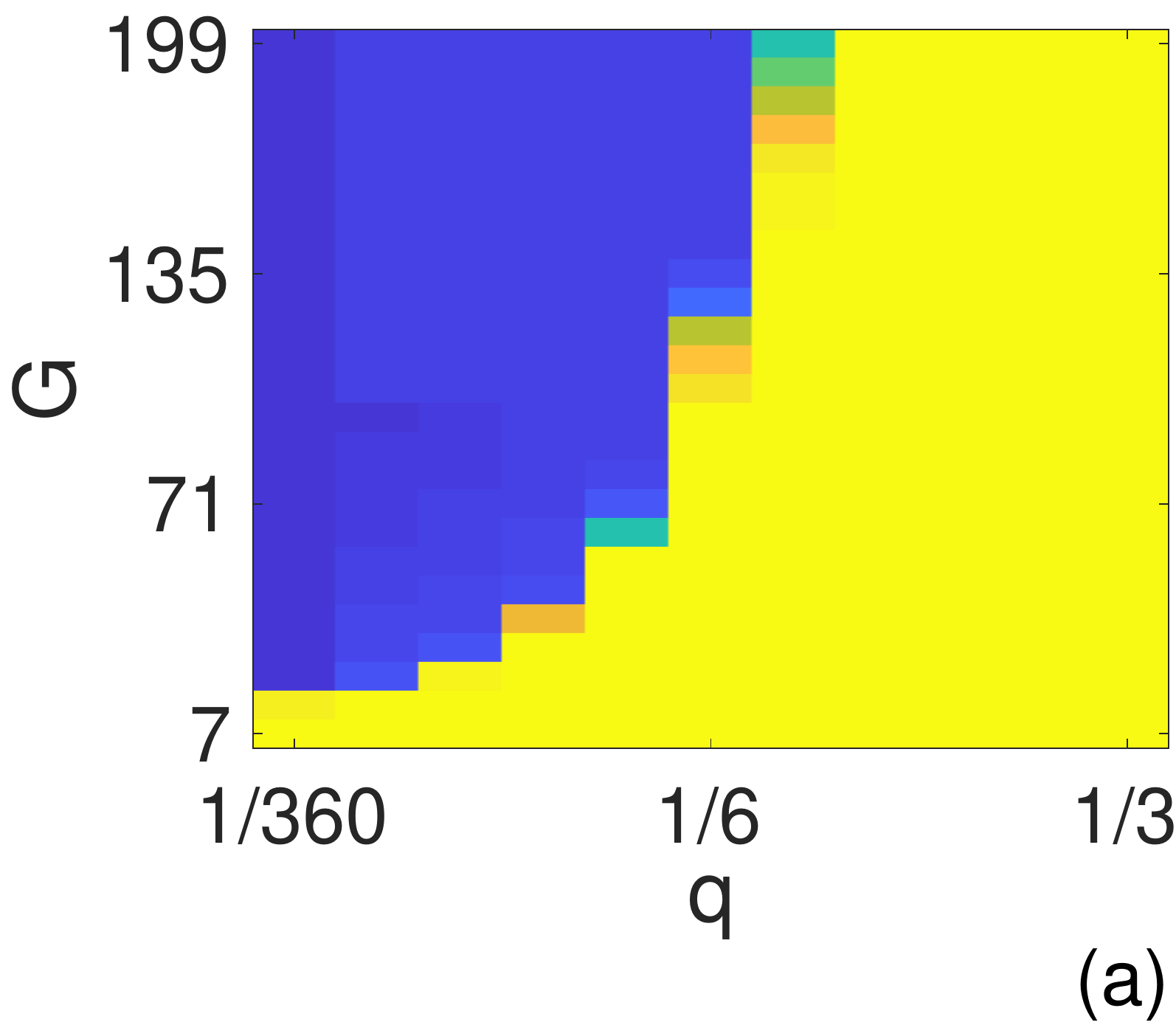}
\includegraphics[scale=0.255]{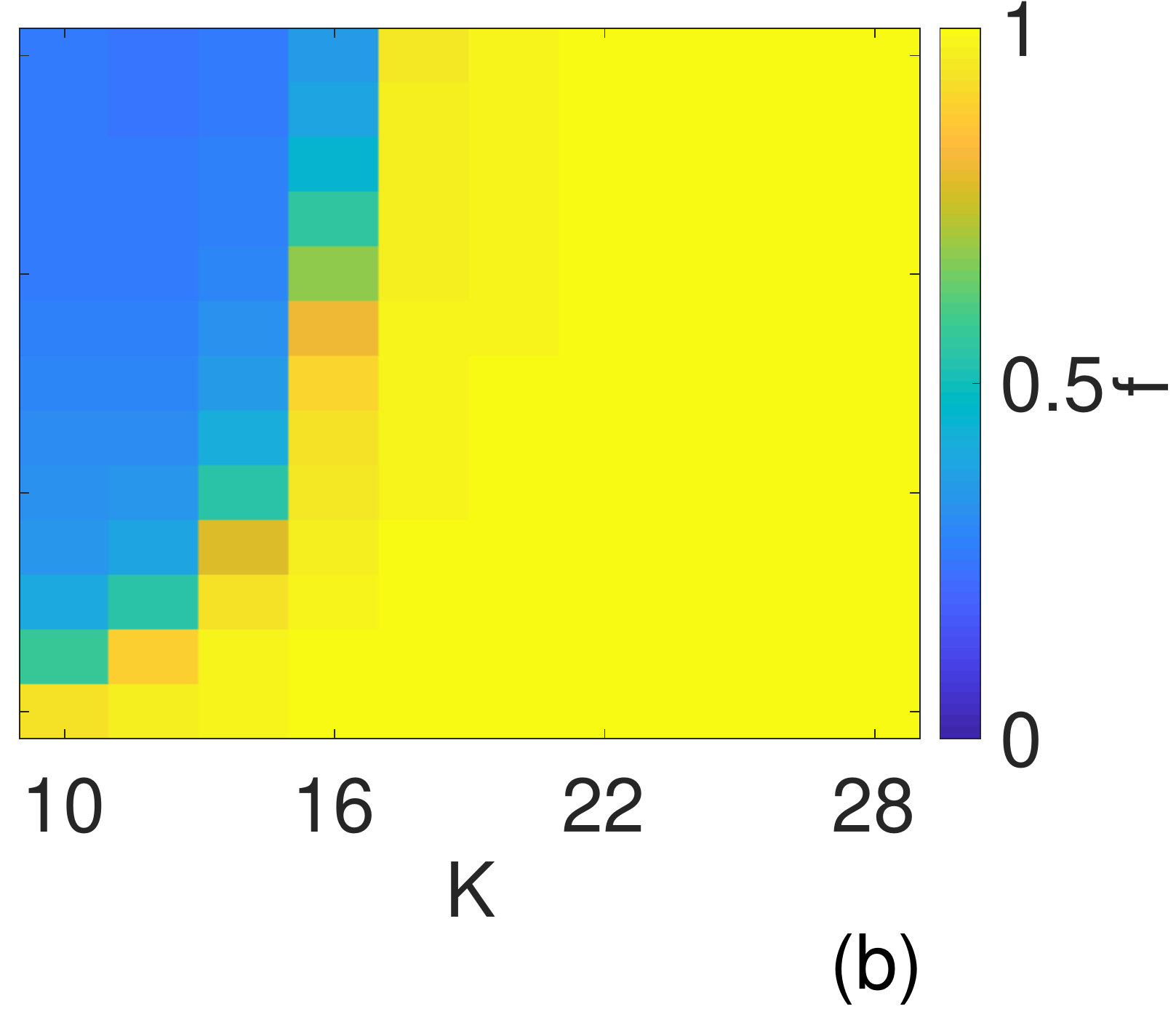}
\caption{Order parameter, $f$, (a) as a function of $q$ and $G$ for $K = 32$, and (b) as a function of $K$ and $G$ with $q = 1/3$. Results are averaged over 100 runs on a regular ring network with $N=400$. 
}
\label{fig:3}
\end{figure}

Next, we run the simulation on a regular ring network for $N=400$, where every node has a degree of $K = 32$ neighbors. Figure \ref{fig:3} (a) shows a phase diagram of the order parameter, $f$, which indicates a transition from an unbalanced (cohesive) to a balanced (fragmented) society. For any given $G$, this transition occurs as $q \equiv Q/N^\Delta$ passes a threshold $q_c$. The existence of a critical $q_c$ demonstrates the importance of Heider's balance term in driving a society towards fragmentation: if many triangles are taken into account, society becomes fragmented. For a wide range of $G$, $q > 1/5$ clearly suffices to be in the balanced and fragmented phase. Also, $q_c$ increases with growing $G$, indicating that when more issues become relevant for homophily, the chance for fragmentation lowers.  Note that the situation resembles non-equilibrium in the sense that in the cohesive phase the system only reaches  quasi-stationary \emph{unbalanced} states which, due to fluctuations in  finite-sized systems, eventually become highly \emph{balanced} after a very long time. Since the presented model is stochastic,  these final states are not necessarily frozen (absorbing). 
This means that a small number of unbalanced triads  still fluctuate over time. The transition is presumably first-order, as a region of bi-stability is numerically observed where the order parameter can be  $f \sim 0$ or $f \sim 1$. 
We checked the hypothesis that for a fixed $q$, the fragmented phase can be reached if the network degree exceeds a critical value, $K_c$; see Fig \ref{fig:3} (b). The transition becomes sharper at higher $K$.

\paragraph*{Limit of small Q.} 
In the limit $Q =  1$ and $G \rightarrow \infty$, the society is expected to reach an unbalanced stationary state. This state  has similar network observables (order parameter, $f$, and the fraction of positive links, $\rho_+$) as those obtained in the ABLTD model \cite{Gorski}, although it  differs in both, the notion of structural balance and the detailed update dynamics. While we consider two  types of balanced triangles (Heider's definition), \cite{Gorski} considers only those with three positive links. Also, \cite{Gorski}  introduces an exogenous bias for flipping negative ties, $p$. This leads to a  ``paradise state'' (all links positive) for $p \geq 1/2$. There is no such bias in our model and positive links can switch to negative, and vice versa, with the same probability. In the stationary state, the number of positive links is therefore approximately equal to that of negative ones. We can show this in a mean-field approach for fully-connected networks; see Supplemental Material. Our approach is motivated by the observation that the two links of a chosen triad are not likely to be flipped simultaneously, as $G \rightarrow \infty$.   Instead, only one  of them would be flipped at every update. A set of rate equations for  triads of different types then is  derived. Its  steady state  solution is shown to be  $f^{({\rm st})} = 0$ and $\rho_+^{({\rm st})} = 1/2$, in full agreement with \cite{Gorski} for $p = 1/3$, and for an arbitrary number of attributes.

\begin{figure}[tb]
\includegraphics[scale=0.25]{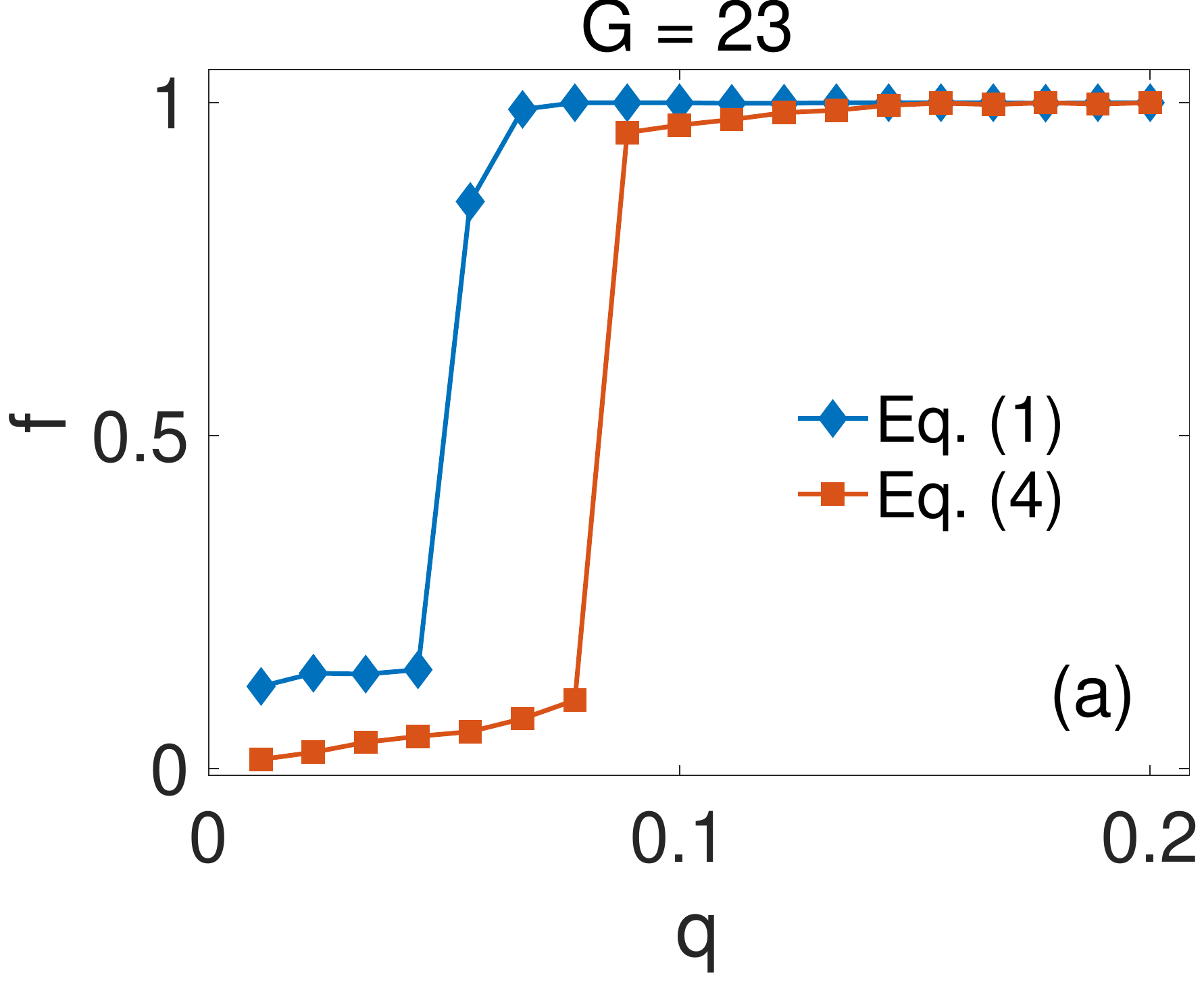}
\includegraphics[scale=0.25]{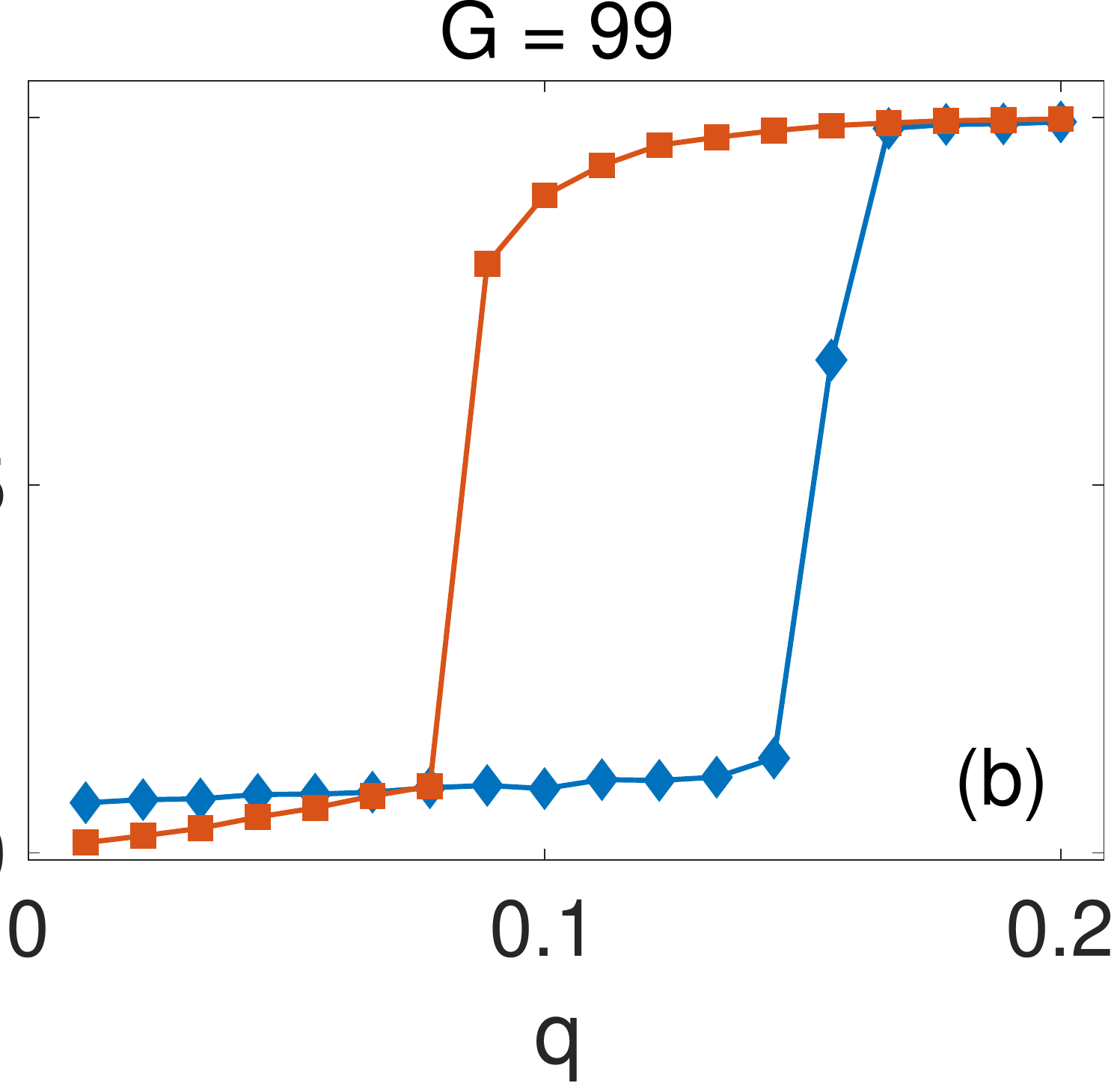}
\caption{Limit of large Q. Comparison of the presented model in Eq. \eqref{local} and the one in Eq. \eqref{global} that resembles \cite{Tuan}. The coupling, $g$, is chosen to be $g = q\equiv Q/N^{\Delta}$, where $Q$ is the number of actually updated triads; $N^{\Delta}$ is the total number of triads. Results are averaged over 100 independent runs for $N=100$, $K = 32$, $N^{\Delta} = 360$, and $G = 23$ (a); $G = 99$ (b). 
}
\label{fig:4}
\end{figure}

\paragraph{Limit of large Q.} Another interesting limit is when $Q \rightarrow N^{\Delta}$. In this case, one can compare the model  with the  Hamiltonian approach used in  \cite{Tuan}, in which the contribution of \emph{all} $N \times N^{\Delta}/3$ triangles, weighted by a coupling $g$, is taken into account:
\begin{equation}
 \bar{H} \equiv  -   \frac{1}{2G}\,  \sum_{(i, j)}  J_{ij}\,  \textbf{A}_i \cdot \textbf{A}_j - g \sum_{(i, j, k)} J_{ij}J_{jk}J_{ki}  \,.
 \label{global}
\end{equation}
Here the first sum extends over all connected pairs, the second over all triangles.
In Eq. \eqref{global}, in contrast to the  model presented here, $J_{ij}$ are random dynamical variables that co-evolve with, but are not strictly determined by the opinion vectors. The detailed updating procedure of  \cite{Tuan}, which aims at minimizing $\bar{H}$, is described in the Supplemental Material. Despite the differences in the concrete update dynamics, for a large enough $Q \geq Q_{MF}$,  
 the two models are expected to yield similar results if $g$ is related to $Q$ by $g = \alpha Q / N^{\Delta}$, for some constant $\alpha$. Here the main idea is  that for  sufficiently large $Q$,  individuals' actions have a similar outcome regardless of their knowledge of the total stress $\bar{H}$ in the society. 
 Figure \ref{fig:4} shows the comparison for $\alpha = 1$. The curve of the presented model indeed crosses that of the model given by Eq. \eqref{global}  at  $q_1 \geq Q_{MF}/ N^{\Delta} \simeq 0.133 $ for $G = 23$ in (a), and at $q_2 \geq 1/6$, for $G = 99$ in (b), where the coupling, $g$, is chosen to be equal to $q_1$ in (a) and $q_2$ in (b), respectively.
The transition in both models also shows that the fragmented and balanced phase necessarily exists if the effect of social balance is sufficiently large with respect to that of homophily,  regardless of the stress and the details of the dynamics under consideration.

We showed that society can achieve structurally balanced states if individuals' opinions co-evolve with their social links so as to minimize their {\em individual} social stress.  The two main structure-relevant principles in social dynamics -- homophily and structural balance -- are simultaneously taken into account by the model. The parameter $G$ controls the dimension of the opinion vectors relevant for homophily, and $Q_i$ specifies how many triangles individuals actually consider for their local social balance. The interplay between the two mechanisms results in a nontrivial phase diagram showing an abrupt  change in patterns of social structure. We find two regimes: fragmentation and cohesion. In the former,  society is fragmented into locally cooperative clusters of agents who are  linked positively within  and negatively between clusters. In the latter, globally percolating cooperation is realized by the existence of a large connected component of positively linked agents. The transition between the regimes is numerically observed at a critical fraction of the considered triangles,  $q_c$, illustrating the main message of the paper: If people 
follow too many friends,
 society is more likely to become fragmented. This message is  robust with respect to the change of the network connectivity; for a fixed value of $q$, we see that the higher the degree, the more likely the society fragments.

In the fragmented phase, most of the triads are found to be balanced, which is seemingly in contrast with the message of \cite{Gorski}.  However, there is no contradiction; the two models can be related. If only one triangle is considered, the presented model corresponds to the particular case of \cite{Gorski} where there is no bias toward friendships. For this case both indeed have the same stationary value of network observables. Note that if  an equivalent to their bias term, $p$, is introduced to Eq. \eqref{local} in the form $h (\sum_{ij} 1 - J_{ij})$  with external field strength, $h$, then the ``paradise state'' can be reached for sufficiently large $h$. 

For large $Q$, the model can be related to a recent co-evolutionary model \cite{Tuan} which corresponds to minimizing the Hamiltonian  in Eq. \eqref{global}. By choosing the coupling, $g = Q/N^{\Delta}$, we show that there exists a value 
$Q_{MF}$ beyond which the two approaches produce very similar results. 
A widespread view in the literature is that fragmentation could emerge from collaborative efforts of individuals to minimize overall stress as local minima of  this stress  \cite{AB1993, Rabbani, Belaza2017}. The existence of $Q_{MF}$, however,  suggests that this view is only a good approximation if individuals keep a large fraction of their local triads balanced.
Previous works studying social fragmentation under the joint effects of homophily and social balance have been in only partial  agreement with Heider's theory. The model presented here shows the possibility of social fragmentation being fully consistent with social balance theory.

\begin{acknowledgments}
This work was supported in part by Austrian Science Fund FWF under P 29252 and P 29032, 
and by the Austrian Science Promotion Agency, FFG project under  857136. 
Simulations were carried out in part  at the Vienna Scientific Cluster. 
 AA acknowledges an internship at CSH. 
\end{acknowledgments}

\bibliography{Spinglass}
 
\clearpage
 \appendix
 
 \onecolumngrid
 \section*{Supplemental Material}
 
 \twocolumngrid
 
\subsection*{The steady state of the stress dynamics in the  limit $Q = 1$, $G \rightarrow \infty$ for a fully-connected network}

\begin{center}
\includegraphics[scale=0.2]{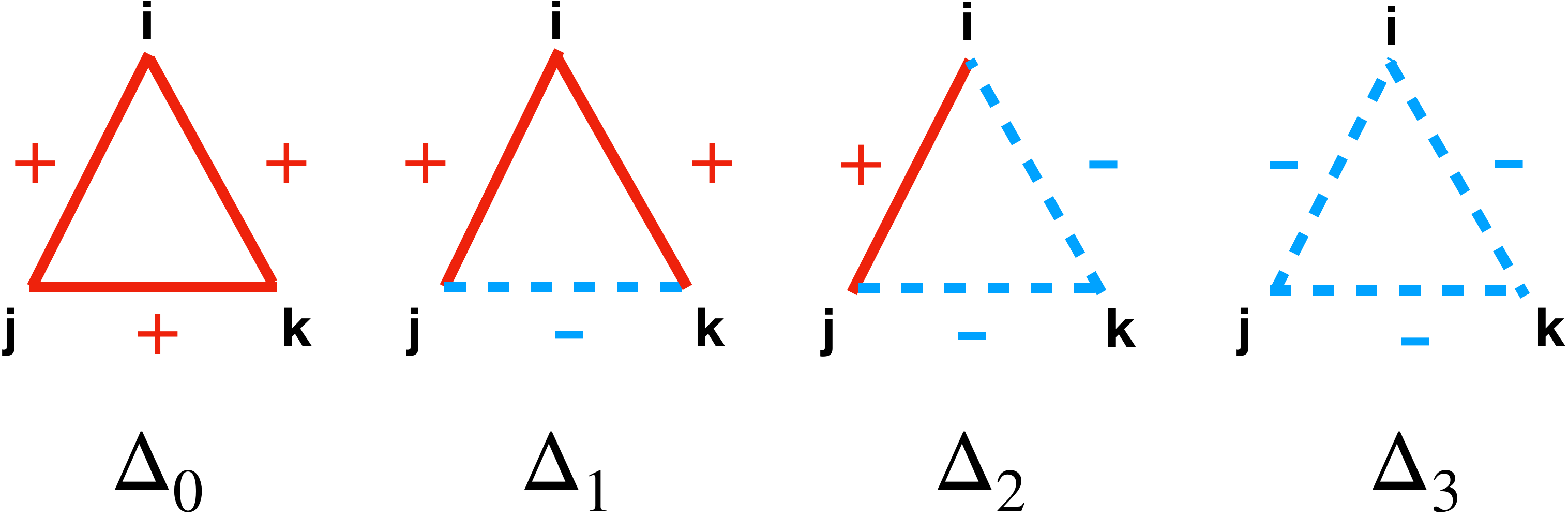}
\end{center}
\textbf{SM Fig. 1.} Four different types of triads: $\Delta_k$, with $k = 0, 1, 2, 3$ being the number of negative links. In the limit $G \rightarrow \infty$, it is very unlikely, for example, that two positive links of the $\Delta_1$ triad attached to $i$ will change their signs at the same time if one of $i$'s attribute flips, resulting in a  $\Delta_3$ triad. 
\vskip0.2cm

When only one triad ($Q = 1$) containing node $i$ is picked in an update,  two links incident to $i$, say $J_{ij}$ and $J_{ik}$,  are to be redefined according to $J_{ij} = {\rm sign}(\textbf{A}_i \cdot \textbf{A}_j)$ and $J_{ik} = {\rm sign}(\textbf{A}_i \cdot \textbf{A}_k)$. From this definition, a change in the link $J_{ij}$ from positive to negative can happen if, and only if, before the update there are $(G+1)/2$ common attributes between $i$ and $j$  and then one among them is selected to flip. This may happen with probability $a _+$. We can estimate 
\begin{equation*} a_+  = \frac{1}{2^{G-1}}\, \binom{G}{(G+1)/2} \times \frac{G+1}{2G}\,,
\label{rate}
\end{equation*} 
assuming $i$ can have any of the $2^{G -1}$ possible states compatible with that of $j$. Similarly,  a previously negative link can become positive with $ a _- = a_+$. This reflects the fact that the process of a link flip is unbiased in our implementation. Note that
\begin{equation*} a_{\pm} \propto \frac{1}{\sqrt{G}}\,, \quad G \rightarrow \infty.
\label{asymptotic_rate}
\end{equation*}
Let  $\rho_+$ be the fraction of positive links. If one assumes that the link $J_{ij}$ can be positive with probability $\rho_+$ or negative with probability $1 - \rho_+$, then  the probabilities that it  changes from $+$ to $-$ and from $-$ to $+$, $\pi^{(+ \rightarrow -)}$ and $\pi^{(- \rightarrow +)}$, respectively,    are given by
\begin{equation*}
\pi^{(+ \rightarrow -)} = \rho_+ a_+ \,, \quad \pi^{(-\rightarrow +)}  = (1 - \rho_+) a_- 
\label{transition}
\end{equation*}
 Among the three possible cases -- namely,  exactly 0, 1, or 2 of the links change sign -- in the  limit $G \rightarrow \infty$, we can neglect the case where both of them change at the same time as this scenario happens with  vanishing probability. This means that in our treatment we consider only those update events during which at most one link can switch. 

We formulate the set of differential equations describing the time evolution of triads in a fully-connected network, following  Antal et al. (Phys. Rev. E72, 036121). Let $\Delta_k$ be as defined in SM Fig. 1. The corresponding density of $\Delta_k$ triads is defined by $n_k = N_k/M$, where $N_k$  is  the total number of triads of type $\Delta_k$ and  $M \equiv \sum_k N_k =  \binom{N}{3}$ is the total number of triads in a fully-connected network. We first define two variables:  the density of triads of type $\Delta_k$ that are attached to a positive link, $n_k^+$, 
$$ n_k^+ = \frac{(3-k) n_k}{3n_0 + 2n_1 + n_2}\,,$$
and the density of triads of type $\Delta_k$ that are attached to a negative link, $n_k^-$,
$$ n_k^- = \frac{kn_k}{ n_1 + 2n_2 + 3n_3}\,.$$
Using $n_k^+$ and $n_k^-$, the set of ODEs can be written as
\begin{equation*} \begin{array}{l} 
       \dot{n}_0   = \pi^{(- \rightarrow +)} n_1^- - \pi^{(+ \rightarrow -)} n_0^+\vspace{0.2cm}\\      \dot{n}_1   = \pi^{(+ \rightarrow -)} n_0^+  + \pi^{(- \rightarrow +)} n_2^- - \pi^{(- \rightarrow +)} n_1^- - \pi^{(+ \rightarrow -)} n_1^+\vspace{0.2cm}\\     \dot{n}_2   = \pi^{(+ \rightarrow -)} n_1^+ + \pi^{(- \rightarrow +)} n_3^-  - \pi^{(- \rightarrow +)} n_2^- - \pi^{(+ \rightarrow -)} n_2^+  \vspace{0.2cm}\\  \dot{n}_3  = \pi^{(+ \rightarrow -)} n_2^+  -\pi^{(- \rightarrow +)} n_3^-.
    \end{array}   \eqno(*) \end{equation*}
The stationary state is defined by setting the left sides of  Eqs. (*)  to zero and requiring $\pi^{(+ \rightarrow -)} = \pi^{(-\rightarrow +)}$ for a fixed density of positive link. This implies
\begin{equation*}
\begin{array}{l}  n_0^+ = n_1^- \,,\quad n_1^+ = n_2^- \,, \quad n_2^+ = n_3^-\vspace{0.2cm}\\   \rho_+^{({\rm st})} a_+ = \big(1 - \rho_+^{({\rm st})}\big) a_-.
\end{array} \eqno(**)  \end{equation*} 
As within  our approximation scheme, $a_+ = a_-$, the steady state  solution to Eqs. (**) is $\rho_+^{({\rm st})} = 1/2$. A consequence of this is the density of triangles type $\Delta_k$ is given by $n_k = \binom{3}{k}/8$, resulting in $f^{({\rm st})} = 0$ in the steady state. These values of $\rho_+^{({\rm st})}$ and $f^{({\rm st})}$ agrees with the finding in (Phys.   Rev.   Lett.   125, 078302) for the case $G < O(N^2)$, $G \rightarrow \infty$   and for $p = 1/3$, $G$ arbitrary, where  $p$ is the tendency towards friendship in their paper. One would need to introduce a $p$ dependence to the expression of $a_+$ and $a_-$ to recover the  upper ($p$-dependent) part of the phase diagram reported there.
 
\section{Effect of the first term in Eq. \eqref{local}}
Here we check what happens to the dynamics if the first term in Eq. \eqref{local} is switched off. We find  in our simulation of finite size systems without the homophily term that   balanced states can still be reached due to large fluctuations in the active phase. These fluctuations, however,   become  smaller as $N \rightarrow \infty$, resulting in a unbalanced quasi-stationary situation. The balanced state hence may not be observed within finite time. This shows that  the effects of both terms are necessary for the convergence towards the balanced state (the sufficient condition is, of course, passing the critical value $Q_c$).
   \begin{center}
   \includegraphics[scale=0.24]{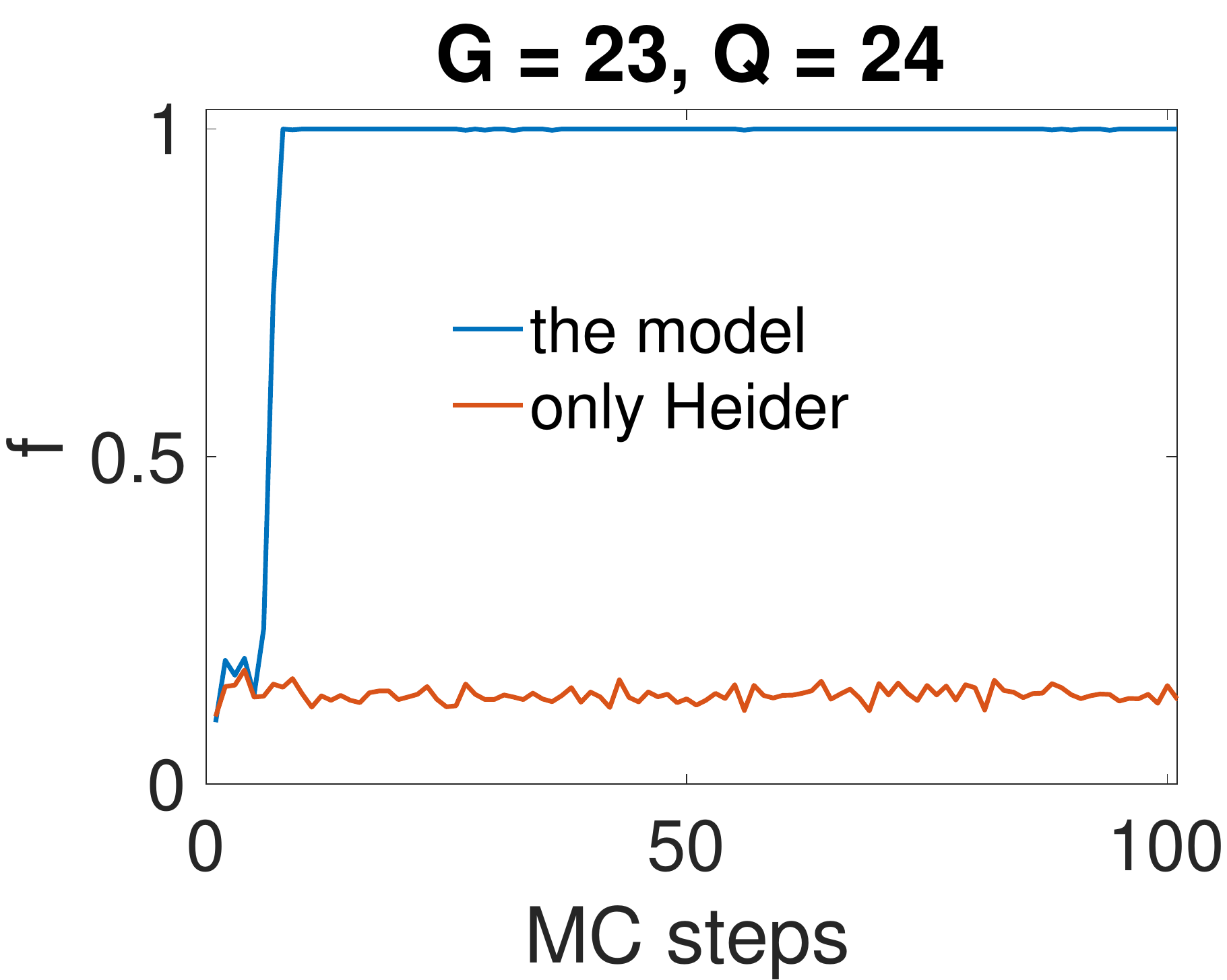}
\includegraphics[scale=0.24]{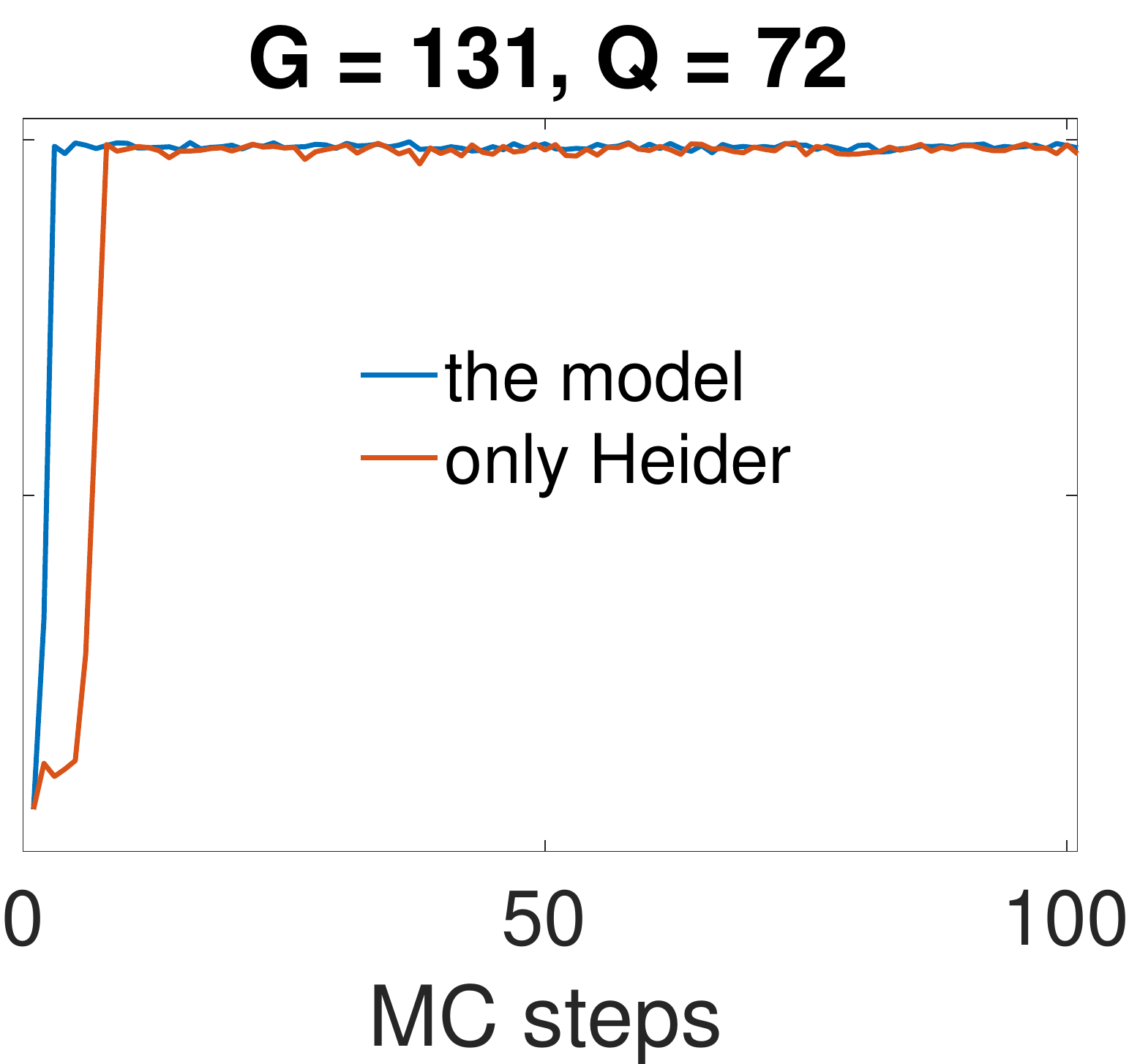}
\end{center}
\textbf{SM Fig. 2.} Timeseries of $f$ for a finite-size system of $N=200$, $G=23, Q = 24$ (left); and $G= 131, Q = 72$ (right) observed in the model simulation on networks with $K = 32$. Here one Monte Carlo step consists of $(N*G)\times(N*K)/2$ opinion flips.

\section{Details of the co-evolution based on the Hamiltonian in Eq. \eqref{global}}

Starting from a random configuration of opinions and links, the society is updated from one timestep $t$ to the next as follows:

\begin{enumerate}
\item Compute $\bar{H}$ of the current state of the system, assume it has a value of $\bar{H}_0$.

\item Pick a node $i$ at random and flip one of its opinions, $a_i^{\ell}$. Compute $\bar{H}$ again, it is now $\bar{H}_1$. 
If the value of $\bar{H}$ has decreased in response to the flip, $\bar{H}_1 \leq \bar{H}_0$, accept the flip. 
If the value of $\bar{H}$ increased, accept the flip only with probability $p = e^{  - \Delta \bar{H} } $, where  
$\Delta \bar{H}= \bar{H}_1- \bar{H}_0$ is the difference of stress before and after the flip. 
Pick the next node randomly and continue until $N$ opinion updates  have been performed.

\item Compute $\bar{H}$ of the system at this point, assume that it is now $\tilde{H}_0$. 
We now pick one link randomly,  $J_{ij}$,  and flip it. Compute $\bar{H}$ again, and assuming it to be $\tilde{H}_1$, we i
accept the flip if $\tilde{H}_1 \leq \tilde{H}_0$, and accept it with probability $p' = e^{  - \Delta \tilde{H} }$, 
where $\Delta \tilde{H} = \tilde{H}_1 - \tilde{H}_0$, if $\tilde{H}_1 > \tilde{H}_0$. 

\item Continue with the next timestep. 
\end{enumerate}

\section{Co-evolution based on the Hamiltonian in Eq. \eqref{global} with links determined by the opinions}

One may ask what would happen to the co-evolutionary dynamics described in  Eq. \eqref{global}  if we update links after every opinion flip according to $J_{ij} = {\rm sign}(\textbf{A}_i \cdot \textbf{A}_j)$ in step 3. This means links are no longer random variables, but strictly determined by agent opinions. Still,  those proposed changes that decrease the Hamiltonian in Eq. \eqref{global}  are favoured over those that increase it, the same as before.  Such a modified dynamics of  Eq. \eqref{global} is rather similar to the presented model.  Indeed,  this variant can also be obtained if the same updating rule of the presented model with $Q = N^{\Delta}$ and the stress in Eq. \eqref{global} are used instead of  that in Eq. \eqref{local}.  SM Figure 3  shows the phase diagram of this variant of  Eq. \eqref{global}  (a) and its section at  various values of $g$ (b) obtained  for a fully-connected network.
A sharp transition between the cohesive and fragmented phases is  also observed. The result suggests further research on a possible correspondence between the Hamiltonian and  our agent-based approach if the  effect of social balance is sufficiently large.
\begin{center}
\includegraphics[scale=0.21]{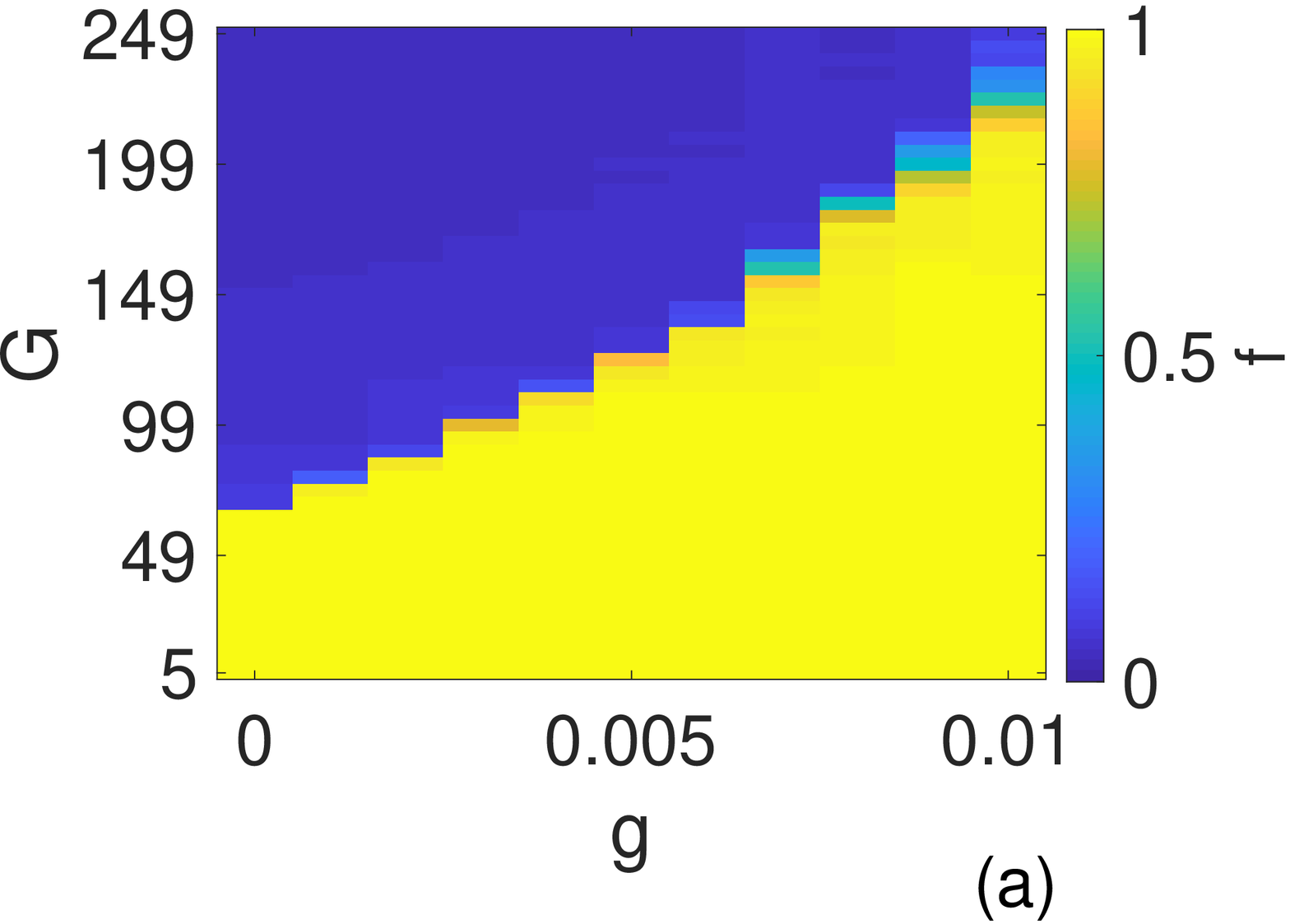}
\includegraphics[scale=0.21]{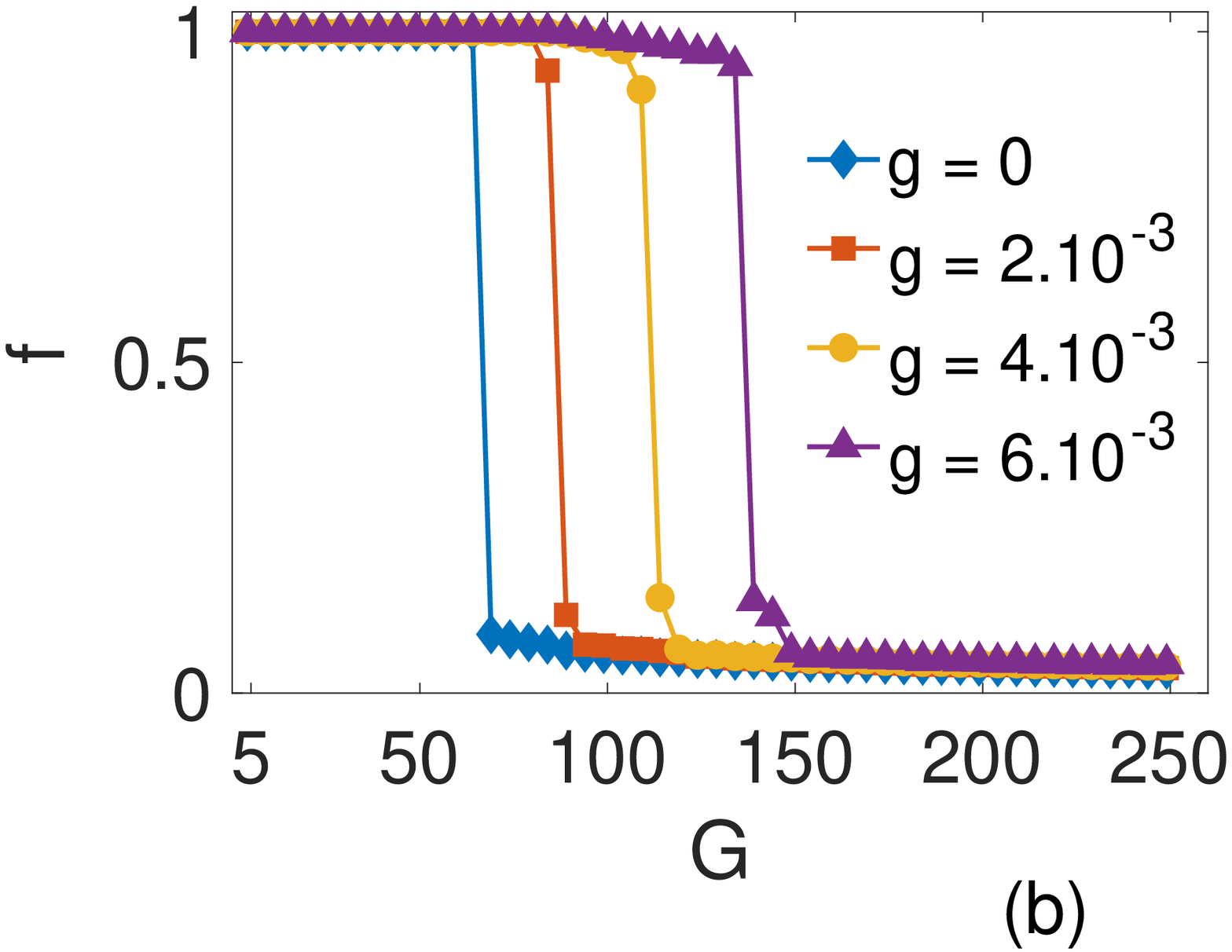}
\end{center}
\textbf{SM Fig. 3.} Result of the model variant of Eq. \eqref{global} with strictly enforcing $J_{ij} = {\rm sign}(\textbf{A}_i \cdot \textbf{A}_j)$. (a) $f$ is shown as a function of  $g$ and $G$. (b) Section of the phase diagram for various strength of the coupling constant, $g$.  $N=100$, results are averaged over 100 runs for fully-connected networks.

  \end{document}